\documentclass[times]{ettauth}
\usepackage{moreverb}
\usepackage{times}
\usepackage{cite}
\usepackage{graphicx,subfigure}
\usepackage[countmax]{subfloat}
\usepackage{color}
\usepackage{amsmath}
\usepackage{amssymb}
\graphicspath{{./figs/}}
\DeclareGraphicsExtensions{{.eps}{.pdf}{.jpeg}}
\newcommand\BibTeX{{\rmfamily B\kern-.05em \textsc{i\kern-.025em b}\kern-.08em
T\kern-.1667em\lower.7ex\hbox{E}\kern-.125emX}}

\def\volumeyear{2010}

\usepackage{setspace}
\begin{document}
\bibliographystyle{plain}
\runningheads{M.~ M.~ Molu, N.~ Goertz}{A Comparison of \journalabb\ class file}

\articletype{RESEARCH ARTICLE}

\title{A Comparison of Soft and Hard Coded Relaying}
\author{Mehdi M.~ Molu\corrauth, Norbert Goertz}
\address{Vienna University of Technology, Gusshausstr.~25/E389, 1040 Vienna, Austria}

\begin{abstract}
``Amplify and Forward'' and ``Decode and Forward'' are the two main
  relaying functions that have been proposed since the advent of
  cooperative communication. ``\textit{Soft} Decode and Forward'' is a
  recently introduced relaying principle that is to combine the
  benefits of the classical two relaying algorithms. In this work, we
  thoroughly investigate \textit{soft} relaying algorithms when
  convolutional or turbo codes are applied. We study the error
  performance of two cooperative scenarios employing soft-relaying. 
   A novel approach, the mutual
  information loss due to data processing, is proposed to analyze the
  relay-based soft encoder. We also introduce a novel approach to
  derive the estimated bit error rate and the equivalent channel SNR
  for the relaying techniques considered in the paper.
\end{abstract}
\maketitle
 \section{Introduction}
Spatial transmit diversity by employing multiple antennas at the
transmitter is one solution to combat fading in wireless
channels. In spite of very promising theoretical results, implementing
multiple antennas in the user nodes can be practically infeasible, if
not impossible, e.g.~due to lack of space. A more recent approach to
exploit spatial diversity is cooperation: several users work together
to communicate with a common destination or even different
destinations, so they can utilize transmit diversity by sharing
resources and obtain better performance, i.e., higher throughput or
lower error rates \cite{SeErAa:2003:I, SeErAa:2003:II,
LaTsWo:2004}.

Two well-known relaying functions
(e.g.~\cite{GoJa:2007}) are "Decode and Forward" (DF) and "Amplify
and Forward" (AF).  A more recent relaying function is ``Soft Decode and Forward''
(soft-DF), e.g.~\cite{SnVa:2005, LiVuWoDo:2006,WeWuKuKa:2008}. The
idea is to combine the benefits of AF and DF and, at the same time, to
mitigate the shortcomings of the traditional algorithms. The core of
soft-DF is a novel Soft-Input Soft-Output (SISO) BCJR encoder that exploits
the trellis structure of the convolutional code. So far, the soft-DF technique has been
evaluated in a scenario where distributed turbo coding
\cite{VaZh:2003} is applied but, interestingly, there is a lack of
literature evaluating recent soft-DF algorithm in simpler scenarios
e.g.~in which the destination simply employs Maximal Ratio Combining
(MRC) instead of advanced iterative decoding algorithms. Studying
soft-DF in different application scenarios reveals more details of the
characteristics of the SISO BCJR encoder used and of soft-DF in
general. Along with SISO BCJR encoder we will introduce a novel
SISO Averaging encoder, which we find has superior
performance compared to the SISO BCJR encoder.

This paper is organized as follows: in Section \ref{Sec:System Model}
we introduce our system model. In Section
\ref{Sec:SoftChannelEncoding} we explain the SISO BCJR and the SISO
Averaging encoders. In Section \ref{Sec:Motivation} we discuss an
observed inconsistency regarding the performance of the soft encoding
algorithms in certain application scenarios. This motivates our
further study of the topic. In Section \ref{Sec:Performance
  Evaluation} we present methods to evaluate the performance of
relaying system using hard and soft channel encoders at the relay. Simulation results
are presented in Section \ref{Sec:Simulation Results}.
In Section \ref{Sec:Discussion} we explain the inconsistency
observed in Section \ref{Sec:Motivation} and section \ref{Sec:Conclusion} offers some conclusion remarks.
\section{System Model}
\label{Sec:System Model}
\begin{figure}[t]  
\begin{center}
 \subfigure[Soft DF]{ \label{fig:SysMod-Soft DF}
\includegraphics[width=0.32\textwidth]{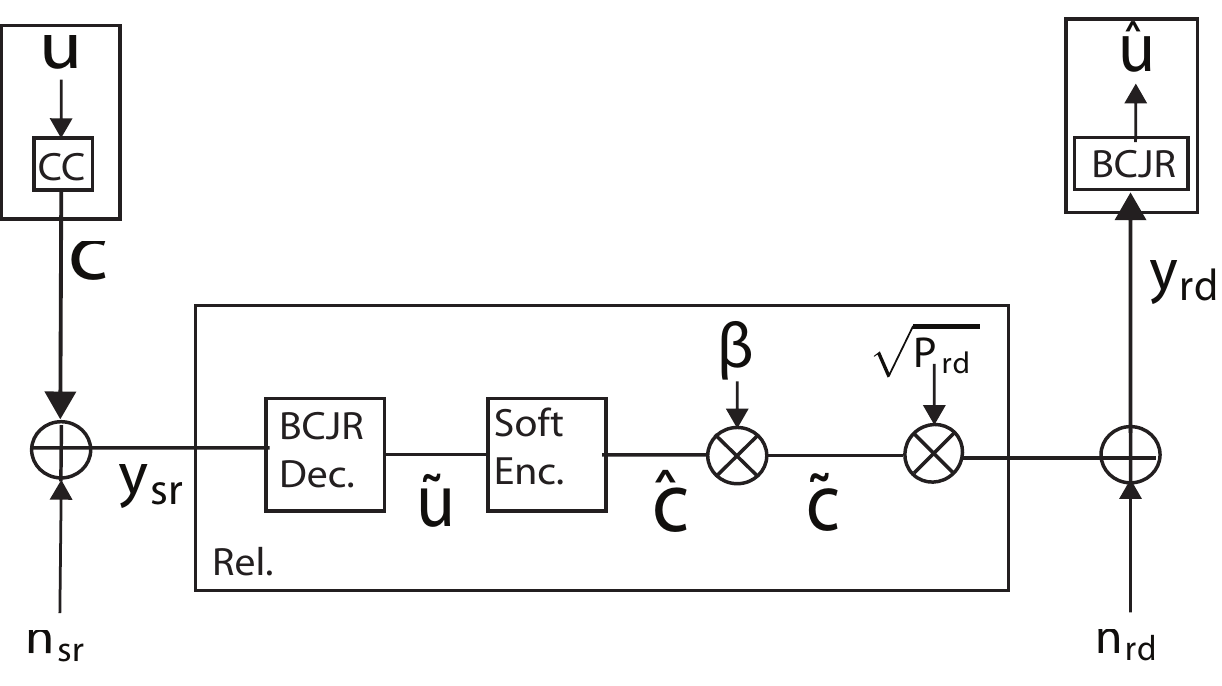}}\\ 
      \subfigure[Soft DTC]{ \label{fig:SysMod-DTC}
           \includegraphics[width=0.36\textwidth]{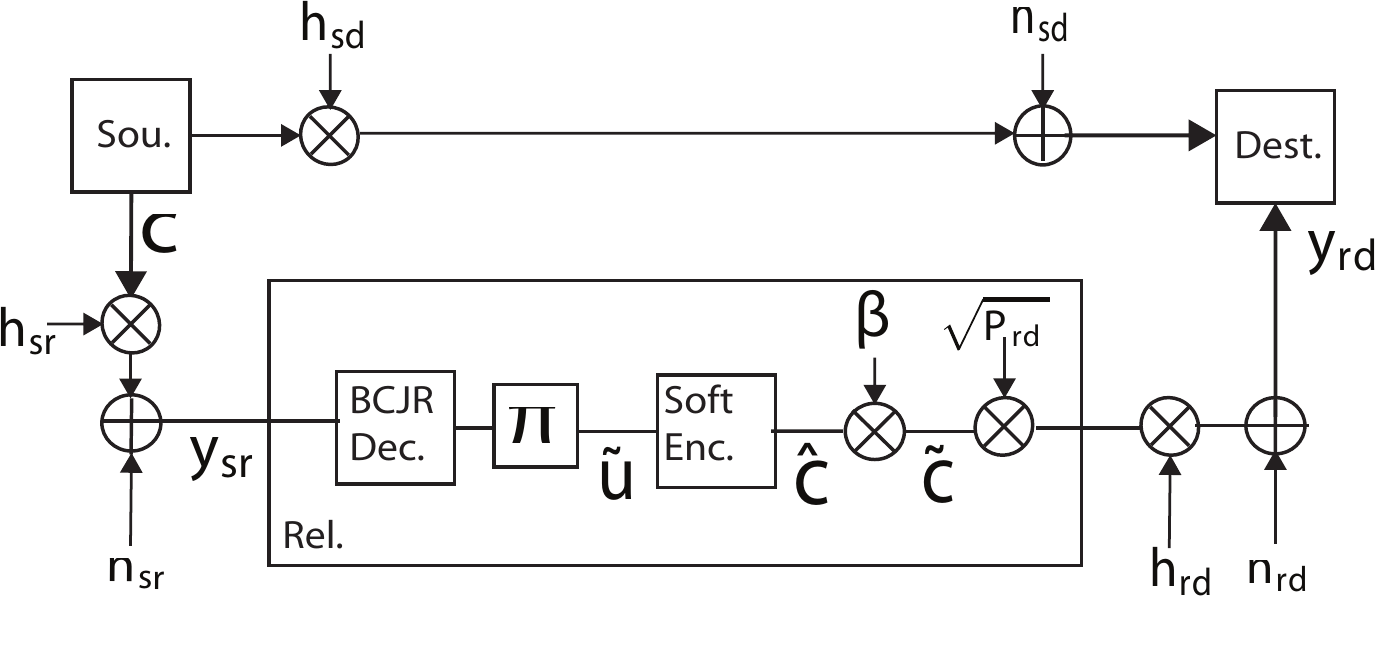}}
\end{center}
\caption{System Model.
 }
 \label{fig:SysMod} 
\end{figure}
We consider a cooperative scenario in which a source node communicates
with a destination via an intermediate relay node. In the sequel, we
introduce two cases of such a cooperative scenario:
\subsection{Case 1}
\label{SubSec: Case1}
Fig.~\ref{fig:SysMod-Soft DF} shows the soft-relaying system under
consideration. We assume that there is \textit{no} direct link between
the source and the destination, which can, e.g., be due to the
distance between the source and the destination.

In the source node, a block of $K$ data bits is encoded using a $k/n$
convolutional encoder, modulated and transmitted towards the
relay. The relay employs a SISO BCJR decoder for decoding the noisy
codeword received via the source-relay link. The output of the SISO BCJR
decoder is fed into a soft channel encoder. The output of the soft channel encoder is scaled by
the factor $\beta$ to fulfil the power constraint of the relay and
transmitted towards the destination. The destination employs the
corresponding BCJR decoder for decoding the noisy codeword received
via the relay-destination link.  

We use two types of soft channel encoders: the SISO BCJR encoder and
the SISO Averaging encoder; both soft encoding algorithms will be
further explained in Section \ref{Sec:SoftChannelEncoding}.

Note that we deliberately make the assumption that the destination
does not ``hear'' the source transmission in order to evaluate
\textit{separately} how the performance is influenced by the soft
information produced by the relay. Thereby, we study the concept of
soft channel encoding as such, without mixing the concept with
 iterative decoding in a distributed Turbo coding scheme
(which will be further discussed in \textit{Case 2}). This is also the
reason why we use a simple convolutional code, as in this case an
optimum symbol-by-symbol decoder (BCJR algorithm) is available
\cite{BCJR:1974}.

\subsection{Case 2} 
\label{SubSec:Case2}
Extending \textit{Case 1} to the more advanced scenario, \textit{Case 2}, we
assume that there is also a direct link available between the source and the
destination. The intention is to construct a Distributed Turbo Code
(DTC) \cite{VaZh:2003} applying soft information relaying
\cite{SnVa:2005,LiVuWoDo:2006,WeWuKuKa:2008}.
Fig.~\ref{fig:SysMod-DTC} shows the soft-relaying system under
consideration. The source functions as in \textit{Case 1}; the
difference is that both the relay and the destination overhear the
data transmitted from the source. Since a Turbo code is applied for
the overall system, we employ an RSC\footnote[1]{Recursive Systematic
Convolutional} encoder in the source. The relay decodes the received
noisy codeword as in \textit{Case 1} and interleaves the
LLR\footnote[2]{Log-likelihood ratio} values prior to soft encoding. The
relay then employs an RSC SISO BCJR encoder for encoding the permuted data
symbols coming out of interleaver. As in any parallel concatenated
Turbo encoder, the systematic bits of the relay encoder are punctured
and only the parity check bits are sent to the destination. Prior to
transmission, power related constraints, as explained in \textit{Case
  1}, are applied to the parity check symbols.
We assume AWGN in the source-relay, the source-destination and the
relay-destination links. For simplicity we employ BPSK modulation in
the source and the relay, but the magnitude of the transmitted
BPSK modulation symbols is weighted according to the power scaling by
the soft-encoded magnitudes of the code bits.

The signals received at the relay ($y_{\text{sr}}$) and the
destination ($y_{\text{sd}}$) at each BPSK symbol time instant,
respectively, are
\begin{equation}
\label{eq:y_sr}
y_{\text{sr}}=\sqrt{P_{\text{s}}} \cdot h_{\text{sr}}
    \cdot c+n_{\text{sr}},\quad c\in \lbrace \pm 1\rbrace
\end{equation}
\begin{equation}
\label{eq:y_sd}
y_{\text{sd}}=\sqrt{P_{\text{s}}} \cdot h_{\text{sd}}
     \cdot c+n_{\text{sd}},\quad c\in \lbrace \pm 1\rbrace
\end{equation} 
and the signal received at the destination equals
\begin{equation}
\label{eq:y_rd}
y_{\text{rd}}=\sqrt{P_{\text{r}}}\cdot\beta \cdot h_{\text{rd}}
   \cdot \hat{c}+n_{\text{rd}},
\end{equation}
where $\beta =1/ \sqrt{\overline{|\hat{c}|^2}}$, with
$\overline{|\hat{c}|^2}$ the average power of the transmitted channel
symbols, averaged over each block of soft-encoded code bits resulting
from each block of $K$ data bits. For simplicity, we assume a non-fading scenario in
\textit{Case 1}, so $h_{\text{sr}}$ and $h_{\text{rd}}$ are unit-power
real values. For \textit{Case 2} we assume Rayleigh fading where
$h_{\text{sr}}$, $h_{\text{sd}}$ and $h_{\text{rd}}$ are zero mean
complex Gaussian random variables each with variance 
$\sigma^2_{\text{h}}$. The noise components, $n_{\text{sr}}$ and
$n_{\text{rd}},$ are zero mean complex Gaussian random variables with
variance $N_0$. The reason for Rayleigh-assumption for
\textit{Case 2} (DTC) will be explained in
Section \ref{Sec:Conclusion}.

There is a competing system design
for the scenarios of \textit{Case 1/2} using hard decisions
for the data bits after soft-input soft-output channel
decoding at the relay, prior to hard re-encoding (by a classical
convolutional encoder), modulation and transmission to the
destination. However, the figures are omitted due to space limits.

The receiver in \textit{Case 1} employs a conventional BCJR decoder
corresponding to the encoder of the relay. The receiver in
\textit{Case 2} applies iterative turbo decoding for the codeword
received partially from the source-destination link and partially from
the relay-destination link.
\section{Soft Channel Encoding}
\label{Sec:SoftChannelEncoding}
In this section we discuss a commonly adopted scheme for soft channel
encoding (Section \ref{Soft-Input Soft-Output BCJR Encoder}) and we
propose a different, much simpler scheme (Section \ref{sec:AvEnc}).

\subsection{BCJR Soft Channel-Encoder}
\label{Soft-Input Soft-Output BCJR Encoder}
The concept of the SISO BCJR \underline{\textit{en}}coder has been
stated in the literature, e.g. \cite{SnVa:2005, LiVuWoDo:2006,
  WeWuKuKa:2008}, but we will also briefly explain it, as we will be
investigating the characteristics of the soft information generated.
We wish to point out here that we follow common practice in the
literature (e.g., \cite{SnVa:2005, LiVuWoDo:2006, WeWuKuKa:2008}) when
we use a \textit{de}coding algorithm (BCJR) for soft channel
\textit{en}coding. The justification is that this soft encoding
algorithm has been reported in the literature to achieve much better
performance in a distributed Turbo coding scheme than hard encoding at
the relay. To the best knowledge of the authors, no theoretical
justification has been given in the literature.

Every BCJR component \cite{BCJR:1974} consists of three main
parameters: $\alpha$ for the forward recursion, $\beta$ for the
backward recursion, and $\gamma$ for the state transition probability.
Assuming an AWGN channel, the BCJR decoder uses the Gaussian
distribution for calculating $\gamma$. However, as explained in
\cite{BCJR:1974}, it is also possible to determine the a-posteriori
probabilities (APPs) of all code bits (not only of the data bits), and
these APPs of the code bits we will consider as the soft
channel-encoder's outputs.

We assume that the outputs of the BCJR decoder (for the source-relay
link) in the relay node are the APPs $P(u_k=\pm 1)$ of the
\textit{data} bits transmitted from the relay, converted to L-value
notation, i.e.
\begin{equation}  
 \tilde{u}_k \doteq \ln \frac{P(u_k=+1)}{P(u_k=-1)} \; .
\end{equation} 
In contrast to transmission over an AWGN channel, for which the BCJR
decoder can use a Gaussian distribution for calculating the transition
probability, $\gamma,$ the SISO BCJR \textit{en}coder
uses the
APPs, $\tilde u_k$, of the data bits, $u_k$, to calculate
$\gamma$. Therefore the parameters $\alpha$, $\beta$ and $\gamma$ for SISO BCJR encoder are
given as
\begin{eqnarray}
\label{eq:AlphaBetaGamma}
&&  \gamma _k(s',s) = \text{P}(u_k,s\mid s') \nonumber\\
&& \alpha_k(s) = \sum_{s'} \alpha_{k-1}(s')\gamma_k(s',s)\\
&& \beta_{k-1}(s') = \sum_{s} \beta_{k}(s)\gamma_k(s',s)  \nonumber  \; ,
\end{eqnarray}
with the summations carried out over all possible states $s',s$ in the
trellis representation of the code and $k$ denoting the time index of
the trellis segment considered (details can be found in
\cite{BCJR:1974}).

The output of the SISO BCJR encoder is the Log Likelihood Ratio
(\textit{LLR} or L-value) $L(c_{k,i} \mid \tilde u_1,\tilde u_2,... )$
of the code bits computed from the input L-values (or corresponding
probabilities $\tilde u_k$) of the data bits:
\begin{equation}
\label{eq:LLR_c}
\centering
L(c_{k,i} \mid \tilde u_1,\tilde u_2,... )  = \text{ln} \frac{\displaystyle\sum_{s'\rightarrow s:\it{c_{k,i}} = 0} \alpha_{k-1}(s')\gamma_k(s',s)\beta_k(s) }  {\displaystyle\sum_{s'\rightarrow s:\it{c_{k,i}} = 1} \alpha_{k-1}(s')\gamma_k(s',s)\beta_k(s) }.
\end{equation}
In (\ref{eq:LLR_c}), the code bit $c_{k,i}$ is the $i$-th code bit of
trellis segment $k$ attached to the data bit input $\tilde u_k$.

The L-values of the code bits derived in (\ref{eq:LLR_c})
are then normalized by $\beta \cdot \sqrt{P_\text{r}} = \sqrt{P_\text{r}}/
\sqrt{\overline{|\hat{c}|^2}}$ across each soft-encoded channel code
word such that the power constraint of the relay is met.

\subsection{Averaging Soft Channel-Encoder}
\label{sec:AvEnc}

The idea of the averaging soft channel encoder is to take the average
of the magnitudes of the L-values of those data bits that are involved
in a parity-check equation that would be used in a classical hard
convolutional encoder. The sign is determined by the normal
parity-check equations, i.e., by ``xor''-operations on the data bits,
with the ``0/1''-output bits mapped to +1/-1 signs for the magnitude
determined above by averaging.

Although such a soft encoder is rather simple it fulfills some
properties that are desirable: if only the signs are considered, the
result will be a valid channel code word. Moreover, it is sensible to
allocate magnitudes to the coded bits that reflect the significances
of the data bits to be encoded by the parity check equations. It
wouldn't make sense to allocate large transmit power to data bits at
the relay when they have been decoded (at the relay) with small
reliabilities. The consequence would be that we communicate to the
destination information that is actually very unreliable, but we would
be making it strong by using large transmit power. On the other hand,
the magnitude should not vanish, when only one of the bits involved in
a parity check has a very small magnitude, as this would cancel the
protection of all other bits as well. The latter point is interesting,
as exactly this will happen when a soft decoding algorithm (such as
the one described in Section \ref{Soft-Input Soft-Output BCJR
  Encoder}) is used as a soft encoder.

We would like to point out that we don't claim the proposed averaging
soft channel encoder to be optimal or even ``good'' in any sense. But,
as we will demonstrate below, the averaging soft channel encoder beats
the SISO BCJR encoder in bit error performance, which proves that this
widely used soft encoder can not be the best choice. 
\section{An Observed Inconsistency}
\label{Sec:Motivation}
Before we continue to study the performance of the two proposed
scenarios, we would like to comment on the motivation of comparing the
two cases. The \textit{Case 2} scenario has been widely studied in the
literature, e.g.~\cite{SnVa:2005,LiVuWoDo:2006,WeWuKuKa:2008}. The
works demonstrate that Distributed Turbo Codes (DTCs) with soft
information relaying outperforms the hard DTC, which our simulations
also confirm (see e.g.~Fig.~\ref{fig:BER_DTC}(solid lines); details of the figure
will be discussed in the forthcoming sections). However, in spite of
the simplicity of the \textit{Case 1} scenario, to the best of our
knowledge, there is as yet no paper considering it.

By resorting to the results of soft-DTC, one may be tempted to
conclude that, in general, a SISO BCJR encoder outperforms
conventional (hard) convolutional encoders.  However, by applying a
SISO BCJR encoder in the \textit{Case 1} scenario we find that hard-DF
achieves better error performance compared to soft-DF (see
e.g.~Fig.~\ref{fig:BER_4dB}). Hence, the results of the two scenarios
\textit{Case 1/2} seem to contradict each other. This unexpected
behaviour of the SISO BCJR encoder in the two proposed scenarios
motivates further analysis. In fact, after observing the results of
the \textit{Case 1} scenario, we can not confirm the strict conclusion
that ``soft coded information relaying is better than hard
relaying''. Therefore, a detailed study of the soft coded information
relaying under different circumstances is provided in this work.
\begin{figure}
\begin{center}
    
        \subfigure[SNR$_{\text{sr}}$ = 0 dB]{
        	                \label{fig:FF_SNR_0}
                        \includegraphics[width=0.21\textwidth]{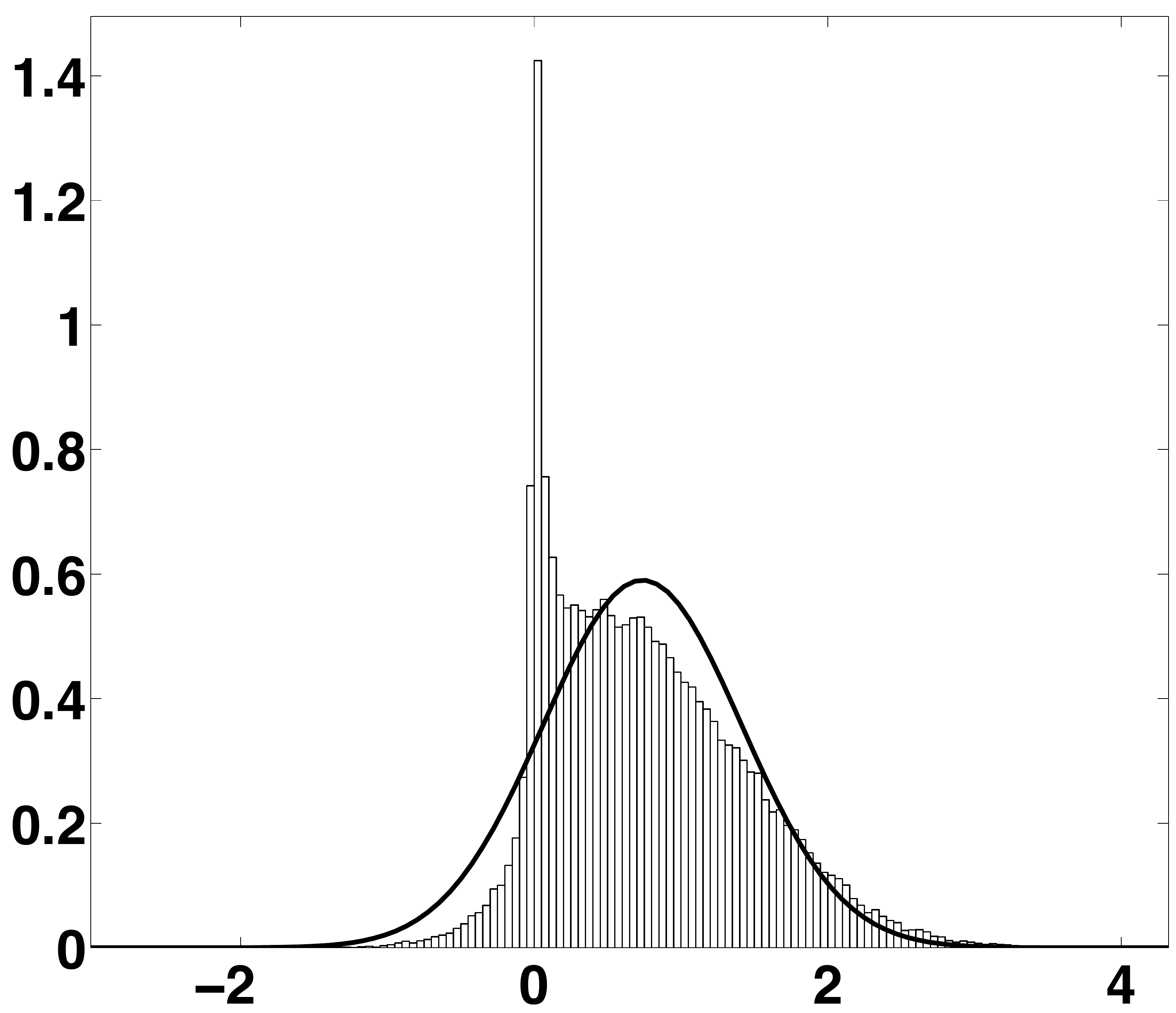}}  
         \subfigure[SNR$_{\text{sr}}$ = 0 dB]{
                       \label{fig:RSC_SNR_0}
                       \includegraphics[width=0.198\textwidth]{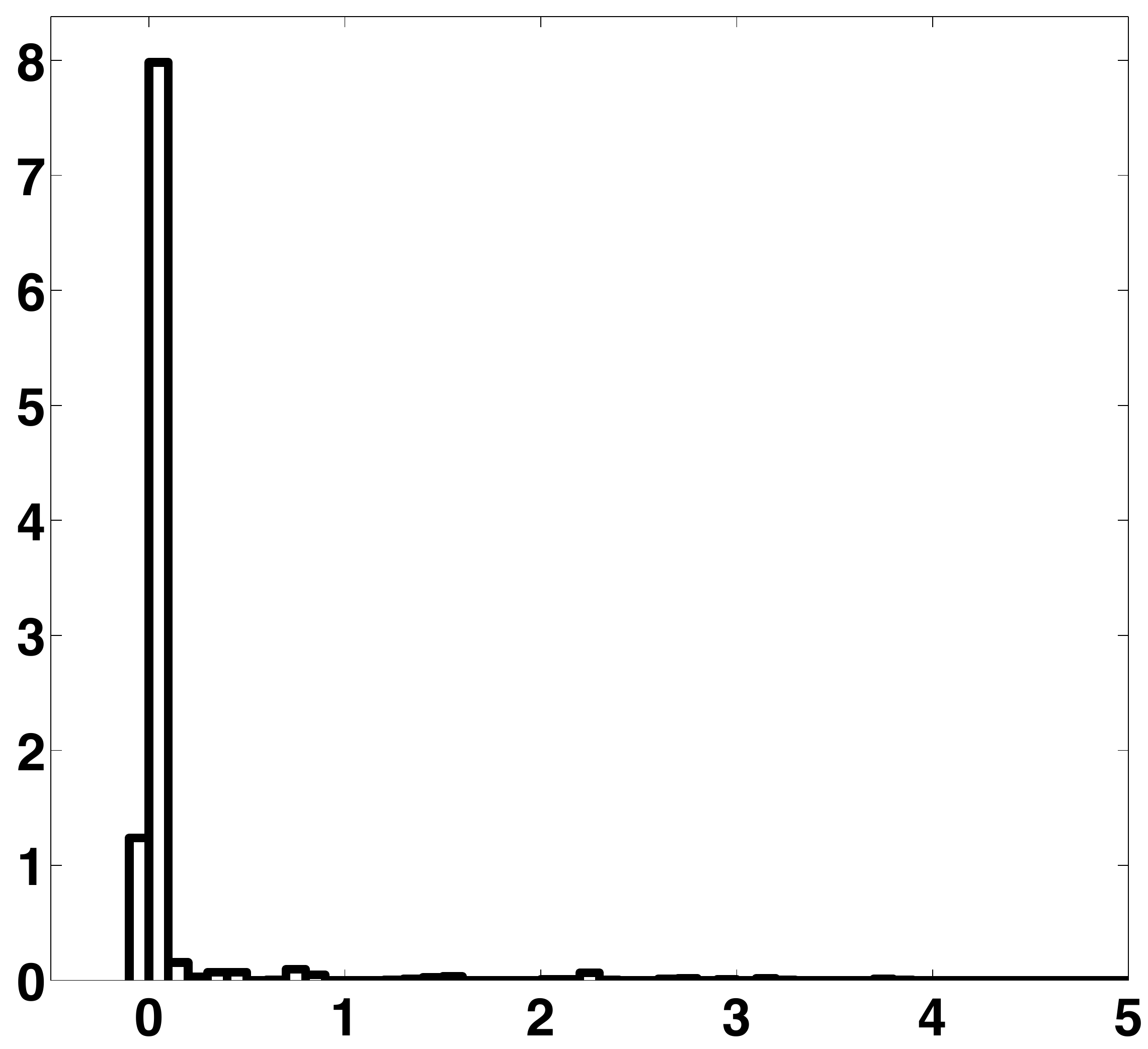}}  
\end{center}
\caption{(a) $p(\tilde c|c =+1)$ for a [7,5] FF SISO BCJR encoder.
(b) $p(\tilde c|c =+1)$ for a [5/7] RSC SISO BCJR encoder. Power normalized to 1. }
\label{fig:Distributions}
\end{figure}
\section{Performance Evaluation}
\label{Sec:Performance Evaluation}
Although turbo codes are appreciated for their
exciting error correction performance in point to point
communications, but in the case of DTCs, however, where two different
nodes (the source and the relay) construct a turbo code, error prone
relays can destroy the performance of the turbo code whenever the relay
forwards erroneous codewords towards the destination. The problem
gets worse during the decoding iterations because of further error propagation
by every iteration. Therefore, we evaluate the performance of SISO BCJR
encoder assuming \textit{Case 1} scenario. The reason is to avoid mixing
the performance of SISO BCJR encoder with the effect of error
propagation per iteration in a turbo decoder of \textit{Case 2}. We are
interested in two parameters: 
\begin{itemize}
\item The mutual information loss due to soft/hard encoding in the relay.
\item The received SNR at destination due to relay transmission.  
\end{itemize}
The statistics of the received signal at the destination corresponding to
the relay transmission depends on the (soft) relaying
function used. To the best of our knowledge, there
is no closed form solution for the probability density function (pdf)
of ${\tilde c}$ for non-trivial relaying functions (e.g. SISO BCJR encoding in the relay). Hence, we have
measured histograms that describe the conditional pdfs $p(\tilde c |
c=1)$. Note that histogram measurement is a common approach used in the literature to
analyse the soft-DF technique.  As an example,
Fig.~\ref{fig:FF_SNR_0} shows the pdf $p(\tilde c | c=1)$ when a
feed-forward BCJR soft channel encoder is applied in the relay,
whereas Fig.~\ref{fig:RSC_SNR_0} shows the
pdf $p(\tilde c | c=1)$ when an RSC BCJR soft channel encoder is
applied in the relay; the pdfs $p(\tilde c | c =-1)$ would be symmetric.
In the literature (e.g.~\cite{SnVa:2005,WeWuKuKa:2008})
the pdfs are usually modeled by
additive zero mean Gaussian random variables, $n_{\tilde{c}}$, plus a
non-zero mean, $\mu_{\tilde{c}}c$, i.e.,
\begin{eqnarray}
\label{eq:c_tilde}
&& \tilde{c} = \mu_{\tilde{c}}c +n_{\tilde{c}}, \: n_{\tilde{c}} \sim \mathcal{N}(0, \sigma_{\tilde{c}} ^2), c\in\{+1,-1\}
\end{eqnarray}
In the reminder of this section we assume that a FF BCJR
encoder is employed in the relay. Nevertheless, we will apply the RSC
BCJR encoder when considering the \textit{Case 2} scenario in the
forthcoming sections\footnote{We emphasis on SISO BCJR encoder because it is the most widely used soft encoder in the literature for soft information relaying. We are not interested in the performance evaluation of SISO Averaging encoder. The SISO Averaging encoder is a competing SISO encoder to show that SISO BCJR encoder is not the optimal SISO encoder. The BER results of SISO Averaging encoder will appear in section \ref{Sec:Simulation Results}. Yet, as will be explained, both the SISO encoders have inferior performance compared to convolutional encoder}.

\subsection{Mutual Information (Loss)}
\label{Mutual Information Loss}
Mutual information, $I(U;\tilde{U})$, can be used to measure the
amount of the information that soft (or hard) data bits, $\tilde{u}$,
at the relay carry about the data symbols, $u$, transmitted by the
source. The two system
models, \textit{soft/hard} DF, use two different (\textit{soft/hard})
channel encoders. The intention of calculating mutual information is
to measure the mutual information loss, \cite{CoTh:2006}, due to
different channel encoders.

The mutual information $I(U;\tilde{U})$ \cite{CoTh:2006,Br:2001}
between the (binary) transmitted data bits, $u\in\{+1,-1\}$, and the
L-values $\tilde{{u}}$ (assuming $\tilde{u}$ is Gaussian distributed \cite{Br:2001})
is given by
\begin{eqnarray}
\label{eq:KuLi}
 I(U; \tilde{U}) &=&\frac{1}{2} \sum_{u' =\pm{1}} \int_{-\infty}^{+\infty} p(\tilde{u}\mid u= u') \\
&&\hspace*{-5mm}\times \log_2 \frac  {2 p(\tilde{u}\mid u = u')}{p(\tilde{u}\mid u = +1) + p(\tilde{u}\mid u = -1)} d\tilde u, \nonumber
\end{eqnarray}
with $p(\tilde{u}\mid u)$ the conditional pdf of the L-values at the
relay (see Fig.~\ref{fig:SysMod}) given the input bits $u$. We have
measured this pdf, similarly as the ones for the code bits ($p(\tilde
c | c =1)$), but we have omitted the plots due to lack of space.

To characterize $I(U;\tilde{U})$ associated with hard-DF, we model the
source-channel-relay link as a Binary Symmetric Channel (BSC) in which
the channel input is a data bit. The output bit of the channel is
flipped with probability $q$. Hence, mutual information for such a BSC
is given (e.g.~\cite{CoTh:2006}) by
\begin{equation}
\label{eq:Mutual_inf_hard}
I(U;\tilde{U}) = 1-H_2(q),
\end{equation} 
with $H_2(q) \doteq -q \cdot \log_2(q) - (1-q) \cdot
\log_2(1-q)$ the standard binary entropy function.

 Using the same approach, one can calculate the mutual information
 $I(C;\tilde{C})$, too (note that $\tilde{C}$ is assumed to be
 Gaussian). A comparison of the mutual informations $I(U;\tilde{U})$
 and $I(C;\tilde{C})$ for both the \textit{hard/soft} DF is useful in
measuring mutual information loss due to employing different encoders in
the relay. Numerical results will follow in section
 \ref{Sec:Simulation Results}.

\subsection{Equivalent Receive SNR at the Destination}
\label{sec:Performance_EQSNR}
In this section we intend to model
the source-relay-destination link with an equivalent AWGN
channel. Note that we employ a convolutional encoder at the source and
a symbol-by-symbol MAP decoder (BCJR decoder) at the destination; the
relay structure has been explained in section \ref{Sec:System Model}.  We
point out that the equivalent receive SNR can be different for another
coding/decoding set up.

\subsubsection{Hard DF}
Due to the error-prone relay, calculating the equivalent receive SNR at the destination for hard DF
is somewhat cumbersome. Since the relay
decodes and forwards both the correct and erroneous frames, the
distribution of the received signal at destination is no longer
Gaussian. The common approach to estimate the SNR at the destination
is to model the source-relay-destination link as an equivalent AWGN
channel with channel $\text{SNR}_\text{eq}$ that depends on both the
source-relay and the relay-destination channel qualities.

The total bit error probability is given by
\begin{eqnarray}
\label{eq:tot_BEP}
P_{\text{tot}}(e\mid \gamma_\text{sr},\gamma_\text{rd}) &=& P_{b}(e\mid \gamma_\text{sr})[1- P_{b}(e\mid \gamma_\text{rd}) ]  \\
 &&+[1- P_{b}(e\mid \gamma_\text{sr}) ] P_{b}(e\mid \gamma_\text{rd}) \nonumber ,
\end{eqnarray}
where $\gamma$ and $P_{b}(e)$ are the corresponding channel SNR and
the bit error probabilities for the two links (source-relay and
relay-destination) involved.

Calculating $P_{\text{tot}}$ using simulations is straightforward but one
can also calculate it using the complementary error
function. The bit error probability of convolutional codes under
symbol-by-symbol MAP decoding (BCJR decoding) can be approximated by
\begin{equation}
\label{eq:erfc}
P_{b}(e) \approx \frac{1}{2} \text{erfc} \Big(  \sqrt{\frac{\mu^2_{\text{out}}}{2\sigma^2_{\text{out}}}} \Big) = \frac{1}{2} \text{erfc} \Big(  \sqrt{ \frac{1}{2}\gamma_{\text{out}}} \Big),
\end{equation} 
(e.g.~\cite{Br:2001}) where $\mu^2_{\text{out}}$ and $\sigma^2_{\text{out}}$ are,
respectively, the mean and variance of the data bit L-values at the
output of the BCJR decoder, and $\gamma_{\text{out}} =\mu^2_{\text{out}}/
\sigma^2_{\text{out}}$.  Similarly, $\mu^2_{\text{in}}$ and $\sigma^2_{\text{in}}$ would
define $\gamma_{\text{in}} = \mu^2_{\text{in}}/\sigma^2_{\text{in}}$ for the input L-values
of the BCJR decoder.
One can model $\gamma_{\text{out}}$ according $\gamma_{\text{in}}$ using
regression analysis; e.g. for a [7,5] BCJR decoder
\begin{eqnarray}
\label{eq:polynomial}
\gamma_{\text{out}}  &=& f(\gamma_{\text{in}})  \\
& \approx &  3.38\cdot10^{-3} \gamma_{\text{in}}^3 -0.12\gamma_{\text{in}}^2 + 2.38 \gamma_{\text{in}} + 2.56. \nonumber
\end{eqnarray}
$ P_{b}(e\mid \gamma_\text{sr})$ and $ P_{b}(e\mid \gamma_\text{rd})$ in
(\ref{eq:tot_BEP}) can be computed using (\ref{eq:erfc}) and
(\ref{eq:polynomial}) (with $\gamma_{\text{in}}
\in\{\gamma_\text{sr},\gamma_\text{rd}\}$). Then, given $P_{\text{tot}}$, the equivalent
SNR$_{\text{out}}$, $\gamma_{\text{eq-out}}$, follows from
  \begin{equation}
  \label{eq:equivalent_SNR}
  \centering
  \gamma_{\text{eq-out}} = 2\Big ( \text{erfc}^{-1}(2P_{\text{tot}})\Big)^2.
  \end{equation}
  By substituting (\ref{eq:equivalent_SNR}) into
\begin{equation}
\label{eq:inv_gama}
\centering
\gamma_{\text{eq}} = f^{-1}(\gamma_{\text{eq-out}}),
\end{equation}
the equivalent source-relay-destination SNR, $\gamma_{\text{eq}}$, is computed.
\subsubsection{Soft DF}
 One might exploit the Gaussian assumption of (\ref{eq:c_tilde}) for
 calculating SNR$_\text{eq}$ of the soft DF schemes, but, as
 illustrated by Fig.~\ref{fig:Distributions}, the Gaussian assumption is
 not accurate, at all, especially at low SNR.  Therefore, in order to
 calculate SNR$_\text{eq}$, we use Monte-Carlo simulations in the
 destination to determine $\gamma_{\text{eq-out}}$. With $\gamma_{\text{eq-out}}$,
 calculating $\gamma_{\text{eq}}$ using (\ref{eq:inv_gama}) is
 straightforward. The estimated BER for soft DF can then be
 determined by substituting $\gamma_{\text{eq-out}}$ in (\ref{eq:erfc}).
\section{Simulation Results}
\label{Sec:Simulation Results}
We start with the simulations for the \textit{Case 1} scenario. Then
we will present the results for \textit{Case 2}, and we will
continue by explaining the reasons why the two scenarios show very
different error performances in spite of employing similar encoders at
the relay. 
 \subsection{Case 1}
The \textit{Case 1} scenario, explained in section \ref{SubSec: Case1}, has the following characteristics: the
source applies a [7,5] convolutional encoder for encoding information
frames of length 2000 bits; for simplicity we use BPSK
modulation.  We assume that
the receive signal at the relay is corrupted by zero mean real Gaussian
receiver noise with a variance of $N_0=1$. 

\begin{figure}[t]
\begin{center}
\includegraphics[scale=0.21]{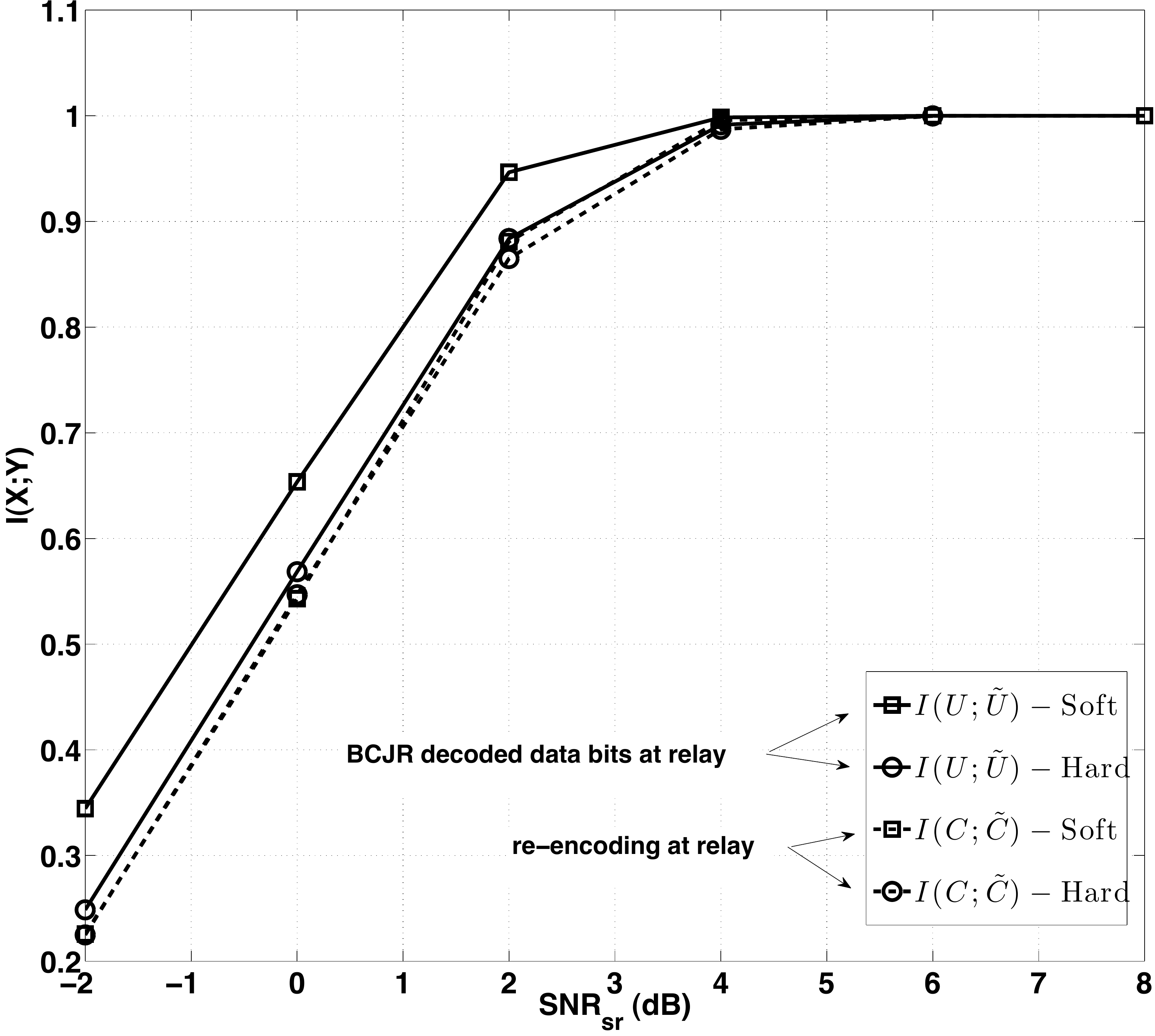}
\end{center}
\caption{Mutual information loss due to hard and soft channel
  encoding. }
\label{fig:mutual_inf}
\end{figure}

Fig.~\ref{fig:mutual_inf} compares the mutual informations $I(U;\tilde
U)$ of the data bits $U$ and their decoded counterparts $\tilde U$
with soft and hard decisions after BCJR decoding at the
relay.  Fig.~\ref{fig:mutual_inf} also shows the mutual informations
$I(C;\tilde C)$ of the code bits $C$ at the source and the (soft and
hard) re-encoded code bits $\tilde C$ at the relay.
The figure shows that the soft channel encoding scheme actually destroys
more information by data processing than the hard encoding algorithm
which has less information at its input. 

\begin{figure}[t]
\centering
\includegraphics[scale=0.235]{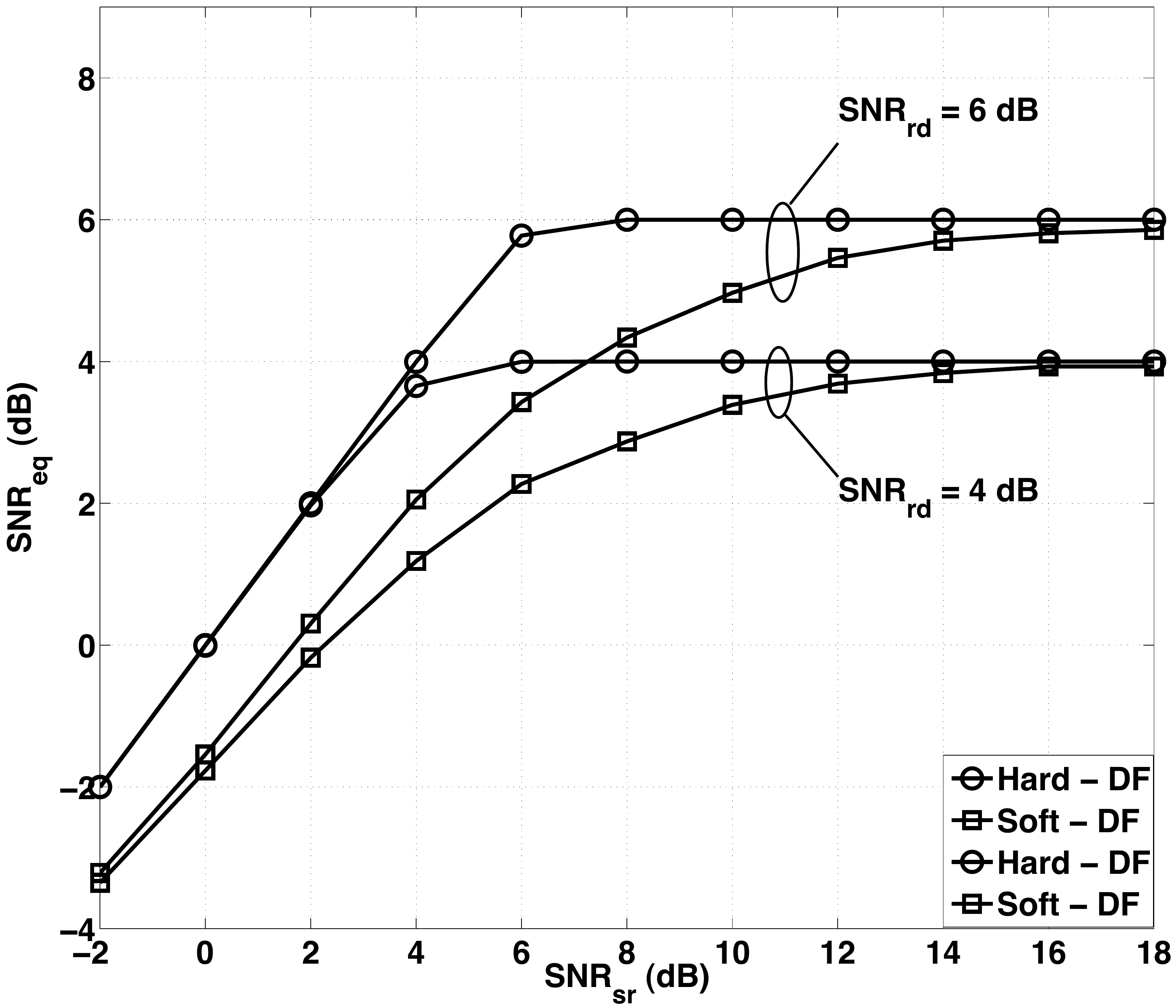}
\caption{Equivalent source-relay-destination SNR; for soft DF, a (7,5)  SISO BCJR channel encoder is used.}
\label{fig:SNRin_SNRout}
\end{figure}

Fig.~\ref{fig:SNRin_SNRout} shows the equivalent
source-relay-destination channel SNR for both the hard DF algorithm
and the soft DF algorithm using a [7,5] BCJR soft channel encoder. The
SNR$_\text{eq}$ plots are calculated according to the rules introduced
in Section \ref{sec:Performance_EQSNR}.
The SNR$_\text{eq}$ curve of soft DF merges with the
SNR$_\text{eq}$ curve of hard-DF at high SNR$_\text{sr}$, although a
small difference remains. The explanation is that the relay usually
performs error free decoding at high SNR; therefore the hard-DF
algorithm is very close to optimum at high SNR. But for soft-DF, the
transmitted symbol from the relay, $\tilde{c}$, is Gaussian
distributed. Because of the unit power constraint of BPSK modulation
that we have to enforce for soft encoding in an average sense as well,
we have an average power of $\text{P}(\tilde{c}) =
\mu^2_{\tilde{c}}+\sigma^2_{\tilde{c}}=1$. Since
$\sigma^2_{\tilde{c}}>0$, we find that $|\mu_{\tilde{c}}|<1$ must
hold, so some of the transmitted code bits will have smaller
instantaneous power than the hard-encoded symbols (which all have
``one''): this will cause the slight SNR$_{\text{eq}}$ degradation in
comparison with hard-DF for large values of SNR$_{\text{sr}}$.

\begin{figure}
\centering
\hspace*{-7mm}
\includegraphics[width=85mm, height=70mm]{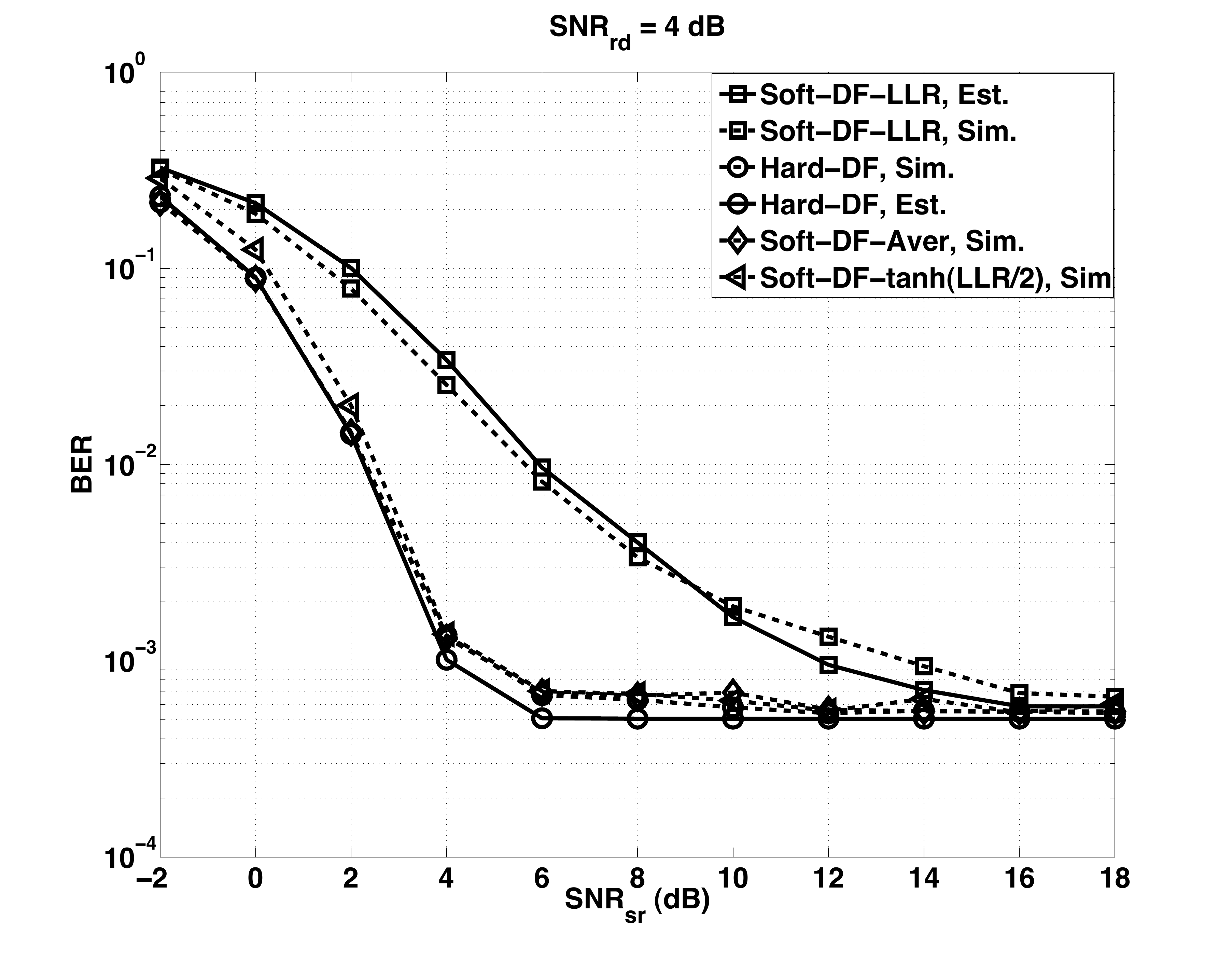}
\caption{BER for the \textit{Case 1} scenario.}
\label{fig:BER_4dB}
\end{figure}

Fig.~\ref{fig:BER_4dB} illustrates the bit error rates of the system
for SNR$_\text{rd}$=4dB. The estimated curves for both the hard and
soft DF (for SISO BCJR encoding) confirm
the simulation results. It is clear that hard DF considerably outperforms 
soft DF when L-values of the codebits are transmitted from the relay as soft information. 
The situation is, however, different for soft DF with the averaging
soft channel encoder: although the performance of hard DF is also
better, it is only slightly so. Of course, this means that hard DF is
still the method of choice, but it also proves that BCJR soft channel
encoding is a much worse algorithm than the averaging soft encoder in
the given context.

In \cite{LiVuWoDo:2006,WeWuKuKa:2008} it has been proposed to perform
a $tanh(\hat{c}/2)$-operation (in Fig.~\ref{fig:SysMod}) after
SISO BCJR encoding and prior to power normalization in the relay. We
did not discuss this in the previous sections. However,
Fig. \ref{fig:BER_4dB} illustrates that even though this modification
outperforms the scenario where L-values of the codebits are
transmitted from the relay, the scheme performs still worse than a
hard convolutional encoder. It is an open question if there is at all
a soft coded DF scheme that can perform better than hard DF within the
frame work of our system model.  
\subsection{Case 2}
The \textit{Case 2} scenario of section \ref{SubSec:Case2} has the following characteristics: the frame
length, modulation and noise characteristics are as in \textit{Case
  1}. A [1, 5/7] RSC convolutional encoder is applied at the
source.  We assume that all the channels $h_\text{sd}$, $h_\text{sr}$ and
$h_\text{rd}$ are subject to Rayleigh fading with unit variance. Fig.~\ref{fig:BER_DTC} shows the BER performance for the \textit{Case
  2} scenario when hard/soft (SISO BCJR) encoding is applied at the
relay and $\text{SNR}_\text{sr}=\text{SNR}_\text{rd} = 12$dB. The figure (solid lines) clearly shows that after
5 iterations, soft relaying outperforms hard relaying with a
considerable difference of about 10 dB when the relay does not perform CRC\footnote[1]{Cyclic Redundancy Check}.
Such a performance behaviour
has been reported in various publications
(e.g.~\cite{SnVa:2005,LiVuWoDo:2006,WeWuKuKa:2008}).
  Apparently, the error
 performance of the \textit{Case 2} scenario contradicts the
 \textit{Case 1} scenario. In \textit{Case 1}, the hard algorithm
 outperforms the soft algorithm while in \textit{Case 2} the soft
 algorithm outperforms the hard algorithm. The open question remaining
 is why the overall system shows such unexpected performance?  This is
 the topic of the rest of the paper.
\section{Discussion}
\label{Sec:Discussion}
\begin{figure}
\centering
\hspace*{-2mm}
\includegraphics[width=75mm, height=70mm]{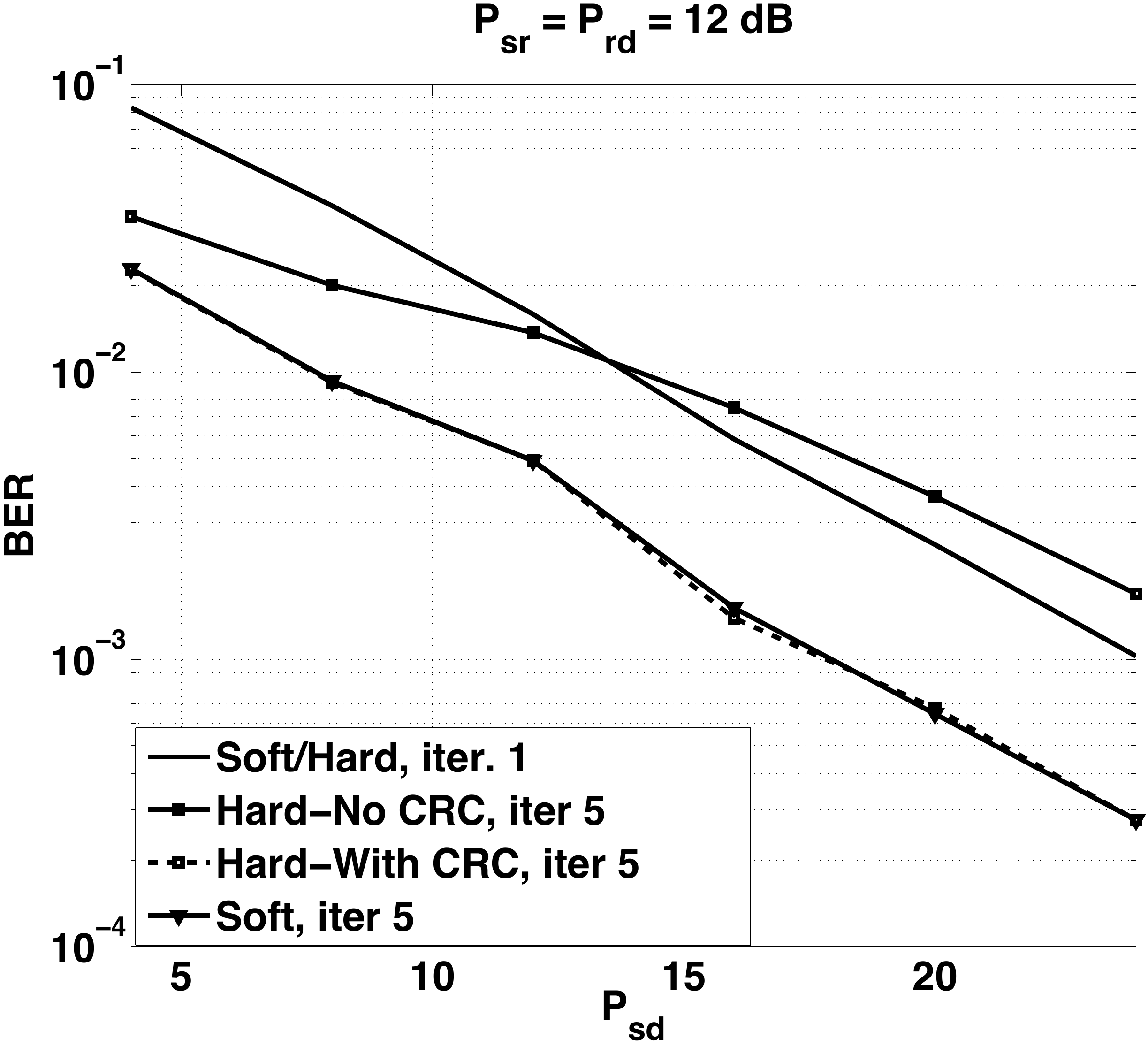}
\caption{BER for the \textit{Case 2} scenario}
\label{fig:BER_DTC}
\end{figure}

\begin{figure}[t]
\begin{center}    
        \subfigure[Y$_{\text{sd}}$]{
        	                \label{fig:Ysd_hard}
                        \includegraphics[width=0.22\textwidth]{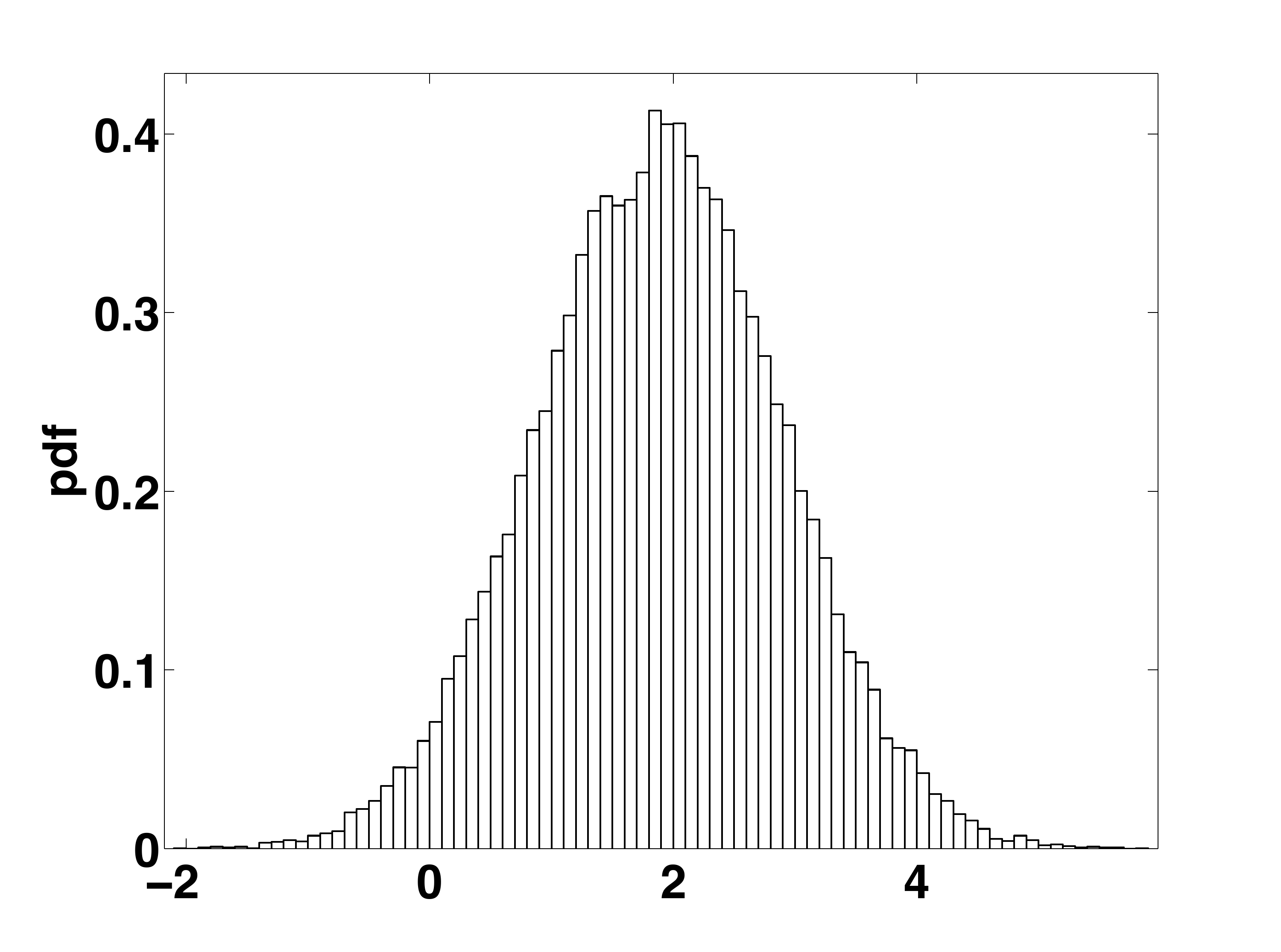}}  
        \subfigure[Y$_{\text{rd}}$]{
                        \label{fig:Yrd_hard}   
                        \includegraphics[width=0.22\textwidth]{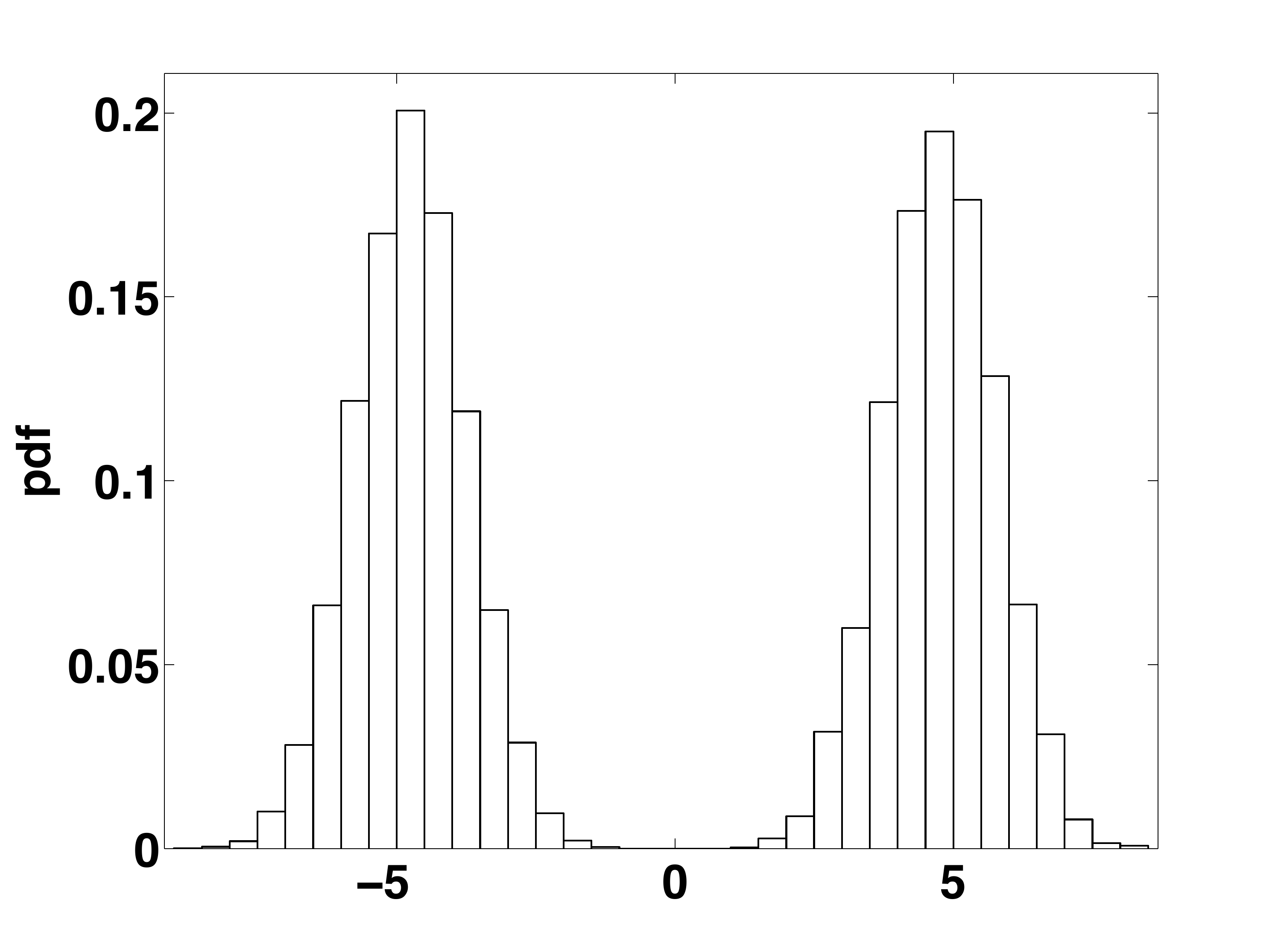}}\\   
        \subfigure[iteration 1]{
                       \label{fig:H1}
                       \includegraphics[width=0.22\textwidth]{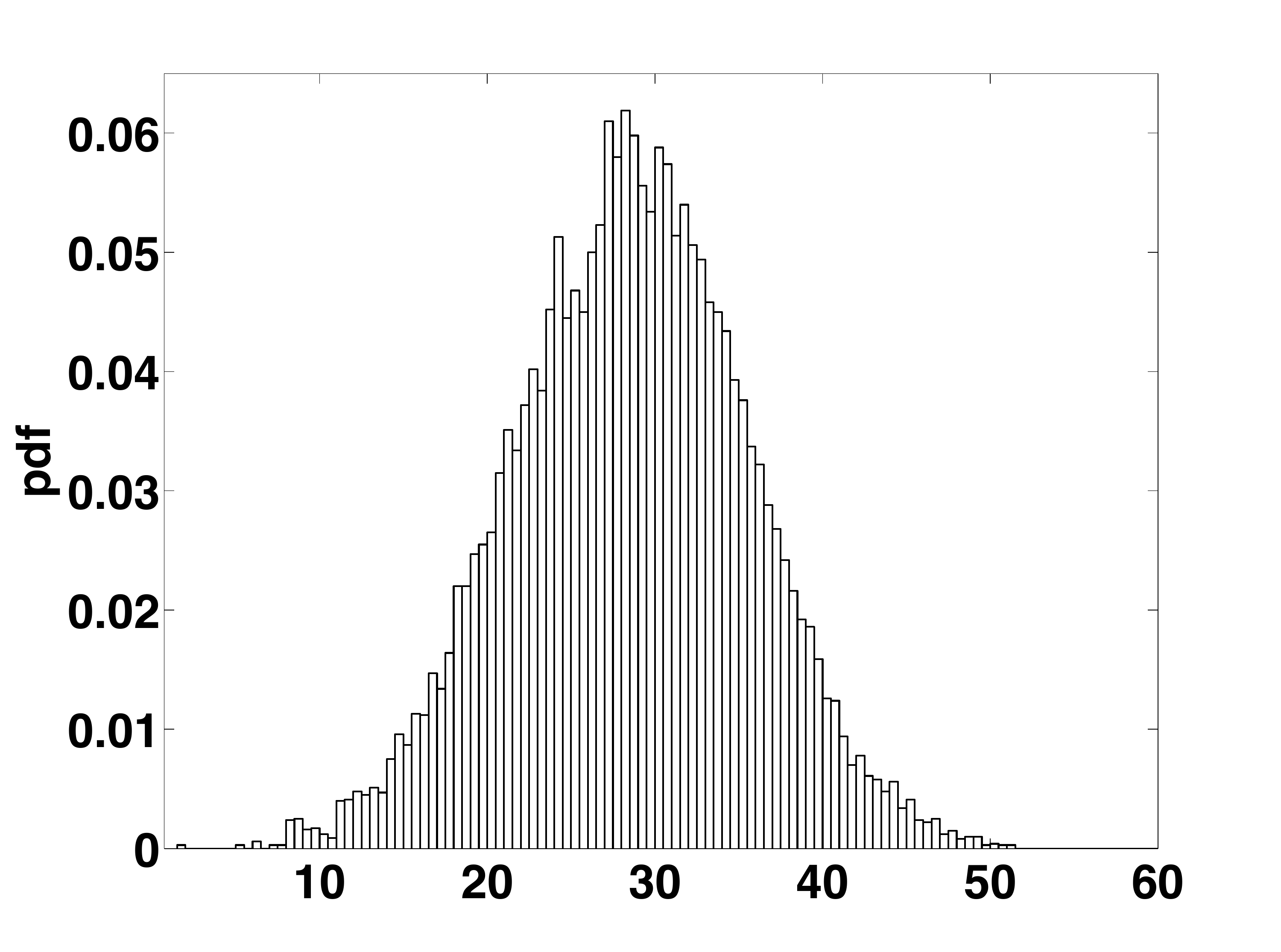}}  
        \subfigure[iteration 2]{
                      \label{fig:H2}
                      \includegraphics[width=0.22\textwidth]{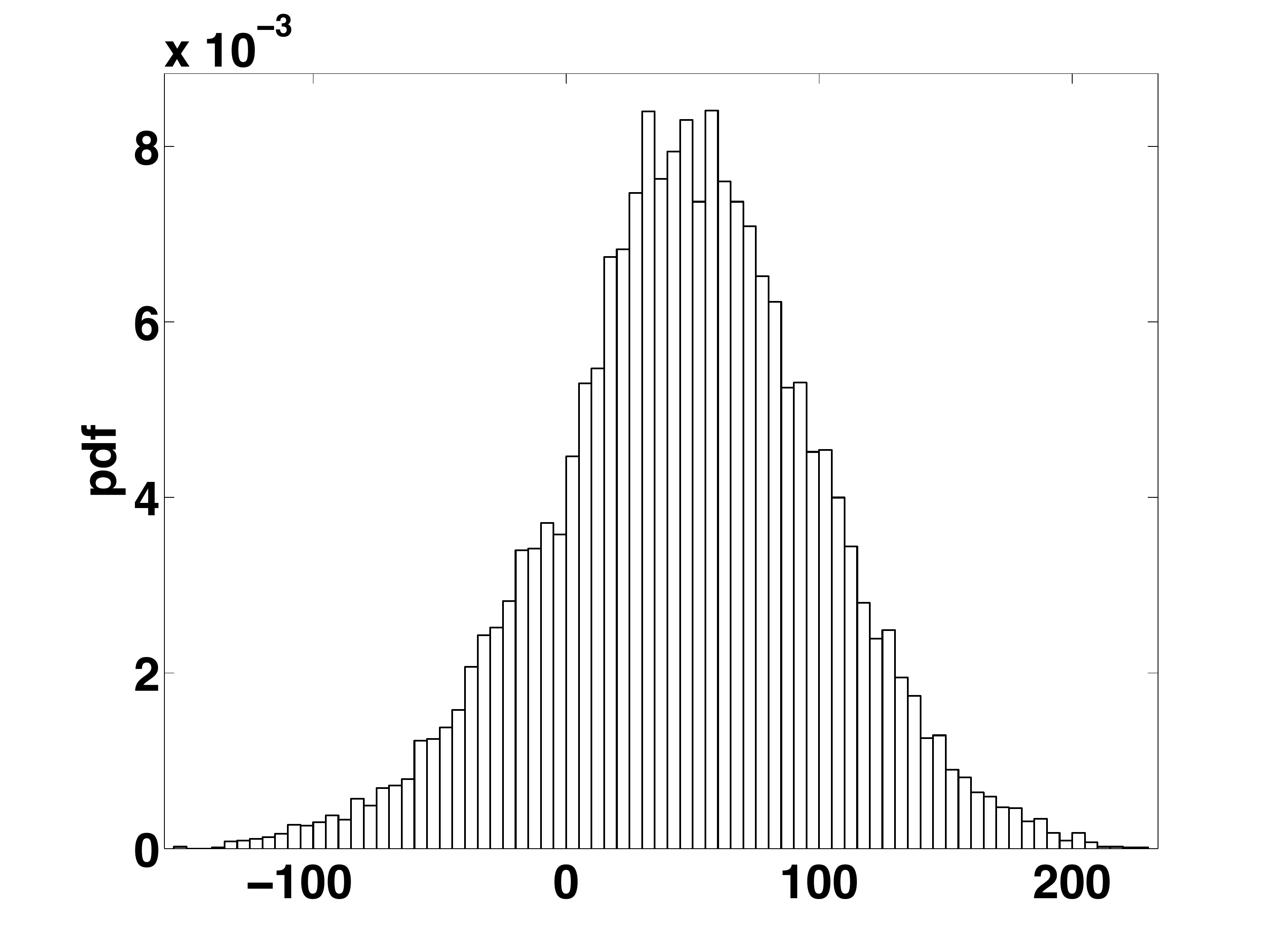}}
\end{center}
\caption{(a) distribution of the received signal at the destination
  with the ''all-zero'' codeword transmitted from the source.
\newline (b) distribution of the received signal at the destination
resulting from  RCS convolutional encoding at the relay.
\newline (c) and (d) distribution $p(L| c= +1)$ of the
information-bit L-values at the destination after the indicated
number of iterations.}
\label{fig:hard_distribution}
\end{figure}

For the analysis of the simulation results of \textit{Case 2} scenario, we assume
that the ``all-zero'' codeword is used. We use a
turbo decoder as the one in \cite{MoGo:2011} where BCJR 1 performs based on the source
transmission whereas  BCJR 2 performs based on both the source
transmission (systematic bits) and the relay transmission (parity check bits).
\paragraph{Hard DTC}
We will first consider the \textit{hard} DTC by considering the
distribution of LLR values of information bits at the output of the
two BCJR decoders after every iteration \footnote[2]{We call one run of each of the BCJR decoders
  ``one iteration''.}. Fig.~\ref{fig:Ysd_hard} shows
the distributions of the received noisy codeword LLRs at the
destination as received from the source; the first BCJR decoder (BCJR 1)
decodes the message given this information and outputs the L-values of the information bits.  
The second BCJR decoder (BCJR 2) uses
three sets of information for decoding: 
\begin{enumerate}
\item  a priori information of the data bits, calculated by BCJR~1
\item receive signal at the destination, transmitted from the
source, (Fig.~\ref{fig:Ysd_hard}), corresponding to the systematic bits of the
codeword 
\item receive signal at the destination, transmitted from the relay,
  (Fig.~\ref{fig:Yrd_hard}), corresponding to the parity-check bits of
  the \textit{supposed} codeword that might be based
  on incorrect decoding at the relay
\end{enumerate}
In the case of decoding failure at the relay, it is unlikely that such a codeword (formed from the systematic bits produced in the source from original data word and parity check bits produced in the relay from an erroneously decoded data word) will exist; in fact, for the all-zero data word
we are sure that it is \textit{not} a valid codeword, because, given all-zero
systematic bits, there is not such a codeword with hamming weight larger than
zero. Therefore, a theoretical
analysis of distributed turbo codes using conventional methods (such as
distance properties) is very difficult (e.g.\cite{TaFoCo:1999,
  BeMo:1996}). That is the reason why we resort to the distribution of the LLR
values for analysis.  
We expect that decoding will fail in decoder BCJR 2
when the relay fails to decode correctly.  Fig.~\ref{fig:H1} shows the
distribution of LLR values of the all-zero databits after the first
iteration. The BCJR 2 component tries to decode databits given
Figs.~\ref{fig:Ysd_hard}, \ref{fig:Yrd_hard} and a priori information
\ref{fig:H1} but it fails to decode correctly because of the problem
described above (non valid codeword). Decoding failure is evident from Fig.~\ref{fig:H2}:
the LLR values of data bits reach values as low as $-100$, when their
signs ``should'' all be positive to be correct. The erroneous a priori
information produced by decoder BCJR 2 will propagate through every
iteration. That is the reason why iteration causes performance degradation, as
illustrated in Fig. \ref{fig:BER_DTC} (solid line). Note that decoding failure in the relay 
 for hard DTC of \textit{Case 2} which employs RSC encoder is much severe than
 decoding failure for
 hard DF of \textit{Case 1} which employs FF encoder. The reason is that even one error bit in the relay encoded by an RSC encoder will propagate through the whole frame; and consequently will cause error burst.
\paragraph{Soft DTC}

\begin{figure}[t]
\begin{center}
    
        \subfigure[Y$_{\text{sd}}$]{
        	                \label{fig:Ysd_soft}
                        \includegraphics[width=0.22\textwidth]{Ysd.pdf}}  
        \subfigure[Y$_{\text{rd}}$]{
                        \label{fig:Yrd_soft}   
                        \includegraphics[width=0.22\textwidth]{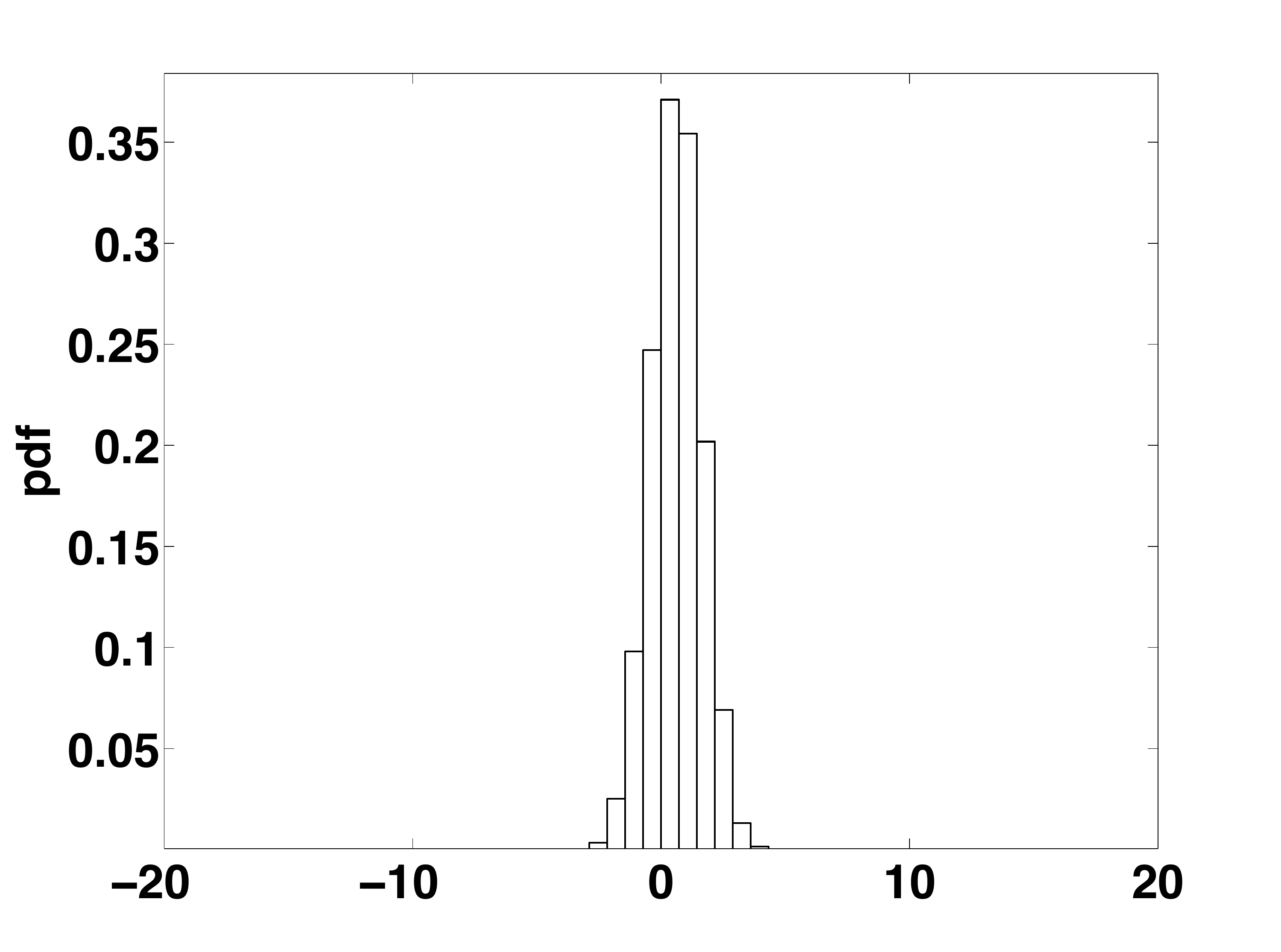}}\\   
        \subfigure[iteration 1]{
                       \label{fig:S1}
                       \includegraphics[width=0.22\textwidth]{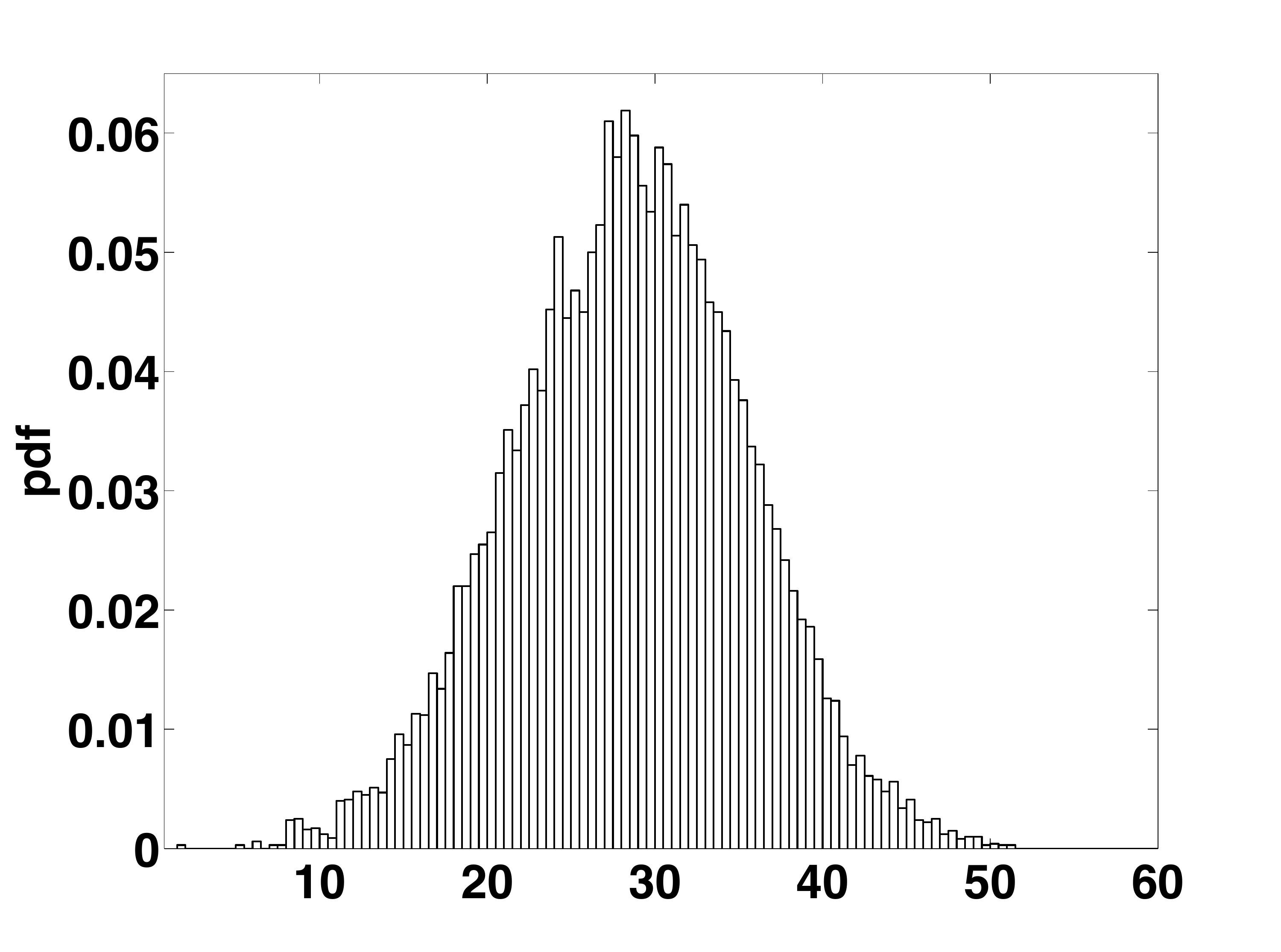}}  
        \subfigure[iteration 2]{
                      \label{fig:S2}
                      \includegraphics[width=0.22\textwidth]{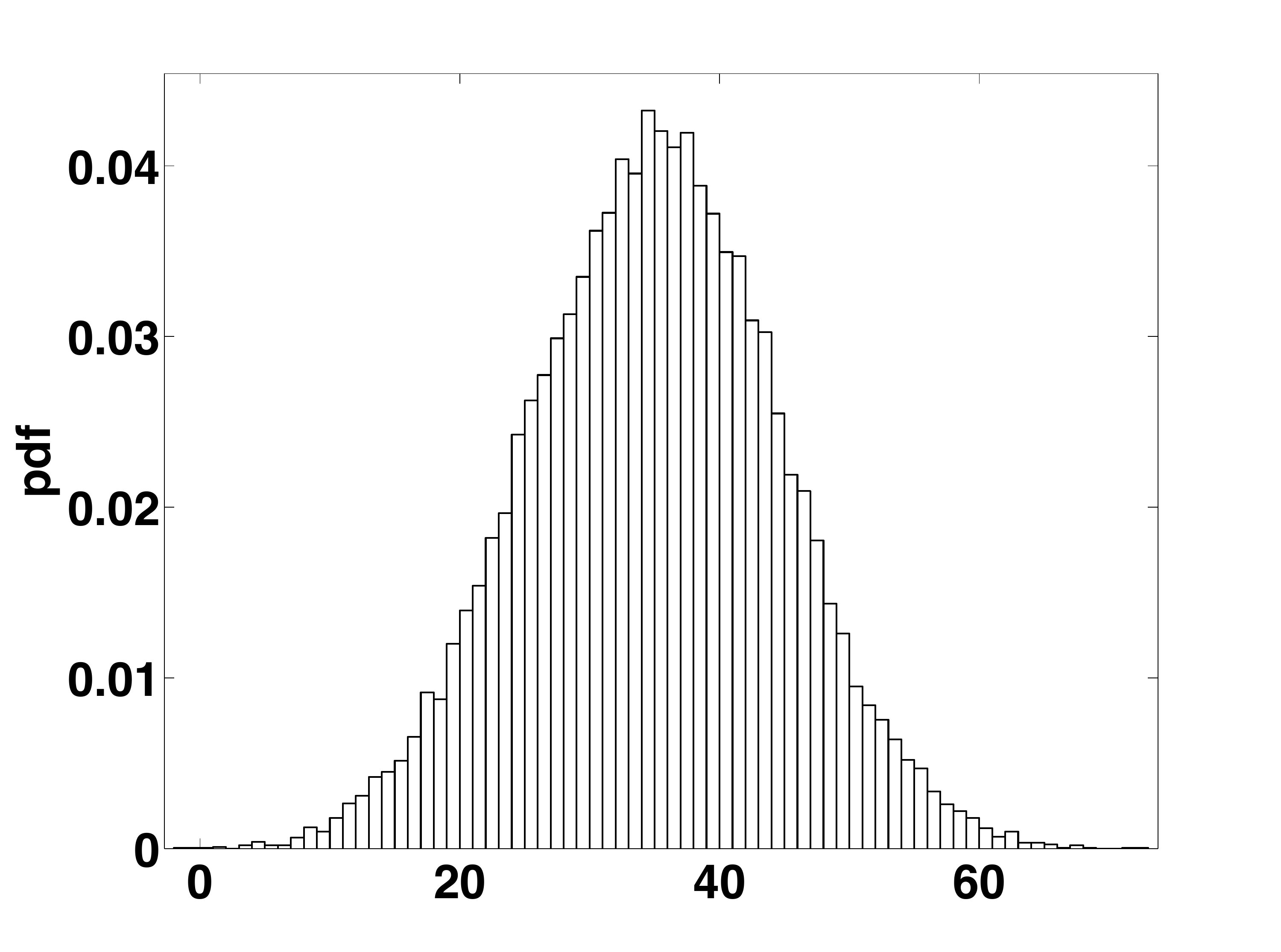}}
\end{center}
\caption{(a) distribution of the received signal at the destination
  with the ''all-zero'' codeword transmitted from the source.
\newline (b) distribution of the received signal at the destination
resulting from  SISO BCJR encoding at the relay.
\newline (c) and (d) distribution $p(L| c= +1)$ of the
information-bit L-values at the destination after the indicated
number of iterations.}
\label{fig:soft_distribution}
\end{figure}

The first iteration of the soft turbo decoder (Fig.~\ref{fig:S1})
works like the first iteration of hard turbo decoder
(Fig.~\ref{fig:H1}). But in the second iteration of the soft turbo
decoder, performance improves -- albeit only slightly -- because the
parity-check bits transmitted from the relay convey some -- albeit
very little -- useful information. But in contrast to hard DTC, in which, 
highly reliable erroneous information (left hand side information 
in Fig. \ref{fig:Yrd_hard} confuse iterative decoder of the destination.
Hence, with soft-DTC there is no such striking performance 
degradation by further iterations as the erroneous information 
from the relay does not appear to be highly reliable. In other words,
since the mean of the transmitted soft signal from the relay tends
to zero in the ``error case'', this set of information will be
treated as noise at the destination. Therefore even though BCJR 2
does not considerably improve the error performance, it also does not
degrade the performance due to incorrect a priori information, unlike
hard-DTC, in which seemingly reliable negative parity-check symbols
confuse decoder BCJR 2.  Figs.~\ref{fig:S2} 
clearly shows that further iterations in the \textit{Case2} scenario 
does not degrade error performance. Nevertheless, further iterations, also,
does not improve the performance.


\paragraph{Implicit CRC by SISO BCJR Encoding}

In this sequel we discuss how a SISO BCJR encoder in the relay
in combination with a BCJR decoder in the destination (as a part of turbo decoder) performs an implicit
CRC. As discussed in previous sections, the
pdf of the codeword L-values at the relay at the output of the SISO
BCJR encoder is assumed to be Gaussian.
The mean ($\mu_{ \tilde{c}}$) of such a distribution
tends to 0, which is clear from Fig.~\ref{fig:RSC_SNR_0}. The BCJR 2 of
the turbo decoder will use the parameters of the assumed distribution
for decoding but since the $\mu_{ \tilde{c}}\to 0$, the relay
bits do not affect the state transition probabilities of BCJR 2. In other
word, since $\mu_{ \tilde{c}}$ of parity
check bits corresponding to the second encoder (SISO BCJR encoder
in the relay) tends to $0$, the BCJR 2 in the turbo decoder at the destination ignores
relay transmission. This property can be interpreted as CRC in the relay for soft
relaying which means that assigning power to such a frame is just a waste of
resources. In fact, better
performance of Soft DTC in comparison with hard DTC is not a consequence of
optimal SISO encoding but a result of error propagation by hard DTC. 
We believe that comparing soft-DTC which implicitly performs
CRC (and, hence avoids error propagation) with hard DTC which is not protected
against error propagation is an unfair comparison.

The arguments of this section are also valid if $\tanh(\hat{c}/2)$
is transmitted instead of the L-values. The reason is that at low
SNR$_\text{sr}$\footnote[1]{Low SNR$_\text{sr}$ is the focus of our
  discussion which occurs repeatedly due to Rayleigh assumption of the channel.}, L-values tend to zero; therefore, $\tanh(\hat{c}/2)$ can be approximated by L-values.
\paragraph{Additional Error Detection}
In order to establish a fair comparison,
 let assume that the relay with hard-DTC performs a
CRC per frame, before transmission. 
Based on that, the relay only transmits the parity check bits if decoding has been
successful; otherwise, the relay
remains silent. 
Fig.~\ref{fig:BER_DTC} (dashed line) shows the BER performance of such a
scenario (CRC-extended ``hard DTC'' and unchanged ``soft DTC'').  The
BER curve of hard-DTC has improved dramatically.  In fact, the  BER
curve of hard-DTC
 performs as it was expected in a turbo decoder: the number of
errors now decreases with each iteration unlike in
Fig.~\ref{fig:BER_DTC} (solid line) in which iterations degraded the hard DTC's
performance. With the mentioned modification, the BER performance of soft and hard DTC are almost equal; but considering the fact that hard DTC saves power in the case of decoding failure in the relay, one can conclude that hard DTC outperforms soft-DTC.
\section{Conclusions}
\label{Sec:Conclusion}
The only advantage of the SISO BCJR encoder appears in the case of decoding
failure at the relay. When the relay fails to decode correctly, error
bursts produced by RSC encoder in hard DTC destroy the performance.  Using a Rayleigh
fading assumption for the channel, we can be sure that, with a certain
(non-zero) probability, there will be channel conditions in which the
source-relay link operates at low SNR and, therefore, error bursts
will indeed frequently destroy the performance of hard-DTC and it is
exactly then when soft-DTC outperforms hard-DTC. Otherwise the SISO
BCJR encoder does not outperform the convolutional encoder in the
sense of mutual information loss nor SNR enhancement. Hence, employing convolutional encoder in the
relay in combination with CRC will be considerably less complex than employing
SISO BCJR encoder which, seemingly, implicit CRC is the only advantage of that.
\bibliography{Ref_Papers.bib}
\end{document}